\documentclass[aps,preprint,10pt]{revtex4}
\usepackage{amsmath}
\usepackage{amsfonts}
\usepackage{amssymb}
\usepackage{graphicx}

\begin{document}

\title{Intermediate-statistics quantum bracket, coherent state, oscillator,
and representation of angular momentum ($su(2)$) algebra}
\author{Yao Shen$^{1)}$}
\author{Wu-Sheng Dai$^{1,2)}$}
\email{daiwusheng@tju.edu.cn}
\author{Mi Xie$^{1,2)}$}
\email{xiemi@tju.edu.cn}
\affiliation{$^{1)}$Department of Physics, Tianjin University, Tianjin 300072, P. R.
China }
\affiliation{$^{2)}$LiuHui Center for Applied Mathematics, Nankai University \& Tianjin
University, Tianjin 300072, P. R. China}
\date{}

\begin{abstract}
In this paper, we first discuss the general properties of an
intermediate-statistics quantum bracket, $\left[ u,v\right] _{n}=uv-e^{i2\pi
/(n+1)}vu$, which corresponds to intermediate statistics in which the
maximum occupation number of one quantum state is an arbitrary integer, $n$.
A further study of the operator realization of intermediate statistics is
given. We construct the intermediate-statistics coherent state. An
intermediate-statistics oscillator is constructed, which returns to bosonic
and fermionic oscillators respectively when $n\rightarrow \infty $ and $n=1$%
. The energy spectrum of such an intermediate-statistics oscillator is
calculated. Finally, we discuss the intermediate-statistics representation
of angular momentum ($su(2)$) algebra. Moreover, a further study of the
operator realization of intermediate statistics is given in the Appendix.
\end{abstract}

\maketitle

\section{Introduction}

Bosons and fermions, the only two kinds of particles that nature realizes,
obey Bose-Einstein and Fermi-Dirac statistics, respectively. For bosons, the
wave function is symmetric and the maximum occupation number is $\infty$;
for fermions, the wave function is anti-symmetric and the maximum occupation
number is $1$. There are two ways to generalize Bose-Einstein and
Fermi-Dirac statistics: (1) One generalization is achieved by generalizing
the symmetry property of the wave function. The wave function will change a
phase factor when two identical particles exchange. The phase factor can be $%
+1$ (symmetric) or $-1$ (anti-symmetric) related to bosons or fermions.
Generalizing this result to an arbitrary phase factor $e^{i\theta}$, one
obtains the concept of anyon \cite{Wilczek}. The corresponding statistics is
fractional statistics. Such a generalization has been applied to the
fractional quantum Hall effect \cite{FH}, high temperature superconductivity
\cite{SC}, supersymmetry \cite{SS}, and quantum computing \cite{QC}. (2)
Another generalization is based on counting the number of many-body quantum
states, i.e., generalizing the Pauli exclusion principle \cite%
{Gentile,Haldane,Wu,HW,Ours1}. A direct generalization is to allow more than
one particle occupying one quantum state. Based on this idea, Gentile
constructed a kind of statistics, called intermediate statistics or Gentile
statistics, in which the maximum number of particles in any quantum state is
neither $1$ nor $\infty$, but equals a finite number $n$ \cite{Gentile}, and
Bose-Einstein or Fermi-Dirac statistics becomes its limiting case when the
maximum occupation number of one state equals $\infty$ or $1$. Many authors
discuss the properties of Gentile's intermediate statistics \cite{P}. Ref.
\cite{Ours2} provides an operator realization of intermediate statistics, by
introducing an intermediate-statistics quantum bracket (a generalized
commutator). In this paper we denote this operation in a more operational
form:%
\begin{equation}
\left[ u,v\right] _{n}\equiv uv-e^{i\theta_{n}}vu,   \label{1}
\end{equation}
where $\theta_{n}=2\pi/(n+1)$ and $n$ is the maximum occupation number of
one quantum state. The bracket, $\left[ u,v\right] _{n}$, will return to
commutativity and anti-commutativity respectively when $n\rightarrow\infty$
and $n=1$:%
\begin{equation*}
\left[ u,v\right] _{n\rightarrow\infty}=\left[ u,v\right] ,\text{ \ }\left[
u,v\right] _{n=1}=\left\{ u,v\right\} .
\end{equation*}
Just as the commutation relation of creation and annihilation operators in
the Bose-Einstein case is commutative and in the Fermi-Dirac case is
anti-commutative, the relation of the creation operator $a^{\dagger}$ and
the annihilation operator $b$ in the intermediate-statistics case obeys an
intermediate commutation relation between commutativity and
anti-commutativity, which, by use of the intermediate-statistics quantum
bracket given in Eq. (\ref{1}), can be expressed as \cite{Ours2}
\begin{equation}
\left[ b,a^{\dagger}\right] _{n}=1.
\end{equation}
Note that creation operator $a^{\dagger}$ is not the Hermitian conjugate of
the annihilation operator $b$ unless $n\rightarrow\infty$ or $n=1$ \cite%
{Ours2}. This realization includes both the phase factor $e^{i\theta_{n}}$
and the maximum occupation number $n$; this implies that such a realization
builds a bridge between the exchange symmetry of identical particles (in
which the phase factor is extended to an arbitrary phase factor $%
e^{i\theta_{n}}$) and the generalized Pauli principle (in which the maximum
occupation number is extended to an arbitrary integer $n$). When the value
of $n$ is given, the phase factor $e^{i\theta_{n}}$ and then the commutation
relation of creation and annihilation operators is determined. In this
scheme different kinds of intermediate statistics correspond to different
commutation relations of creation and annihilation operators. The
commutation relation of creation and annihilation operators of intermediate
statistics is intermediate between commutativity (the Bose-Einstein case),
and anti-commutativity (the Fermi-Dirac case) \cite{Ours2}. In this operator
realization of Gentile statistics, the state $\left\vert \nu\right\rangle
_{n}$ satisfies $a^{\dagger}|n\rangle_{n}=0$ and $b|0\rangle_{n}=0$, where $%
\nu$ is the occupation number and subscript $n$ represents that such an
operator realization corresponds to Gentile statistics with $n$ as its
maxium occupation number \cite{Ours2}.

Although the elementary particles in nature are either bosons or fermions,
the theory of intermediate statistics can be applied to describe
composite-particle systems. Various composite-particle systems have been
studied for many years \cite{Heinonen}, e.g., the Cooper pair in the theory
of superconductivity, the Fermi gas superfluid \cite{OG}, and the exciton
\cite{BGC}, etc. Some composite particles, composed of several fermions, may
behave like bosons, obeying Bose-Einstein statistics, when they are far from
each other. However, when they come closer together, the fermions in
different composite bosons will begin to feel each other, and the statistics
of the composite particles will somewhat deviate from Bose-Einstein
statistics \cite{JGR}. In this case intermediate statistics can be used as
an effective tool for studying such systems.

We will first give a general discussion of the intermediate-statistics
quantum bracket, $\left[ u,v\right] _{n}$, and give a further study of the
operator realization of intermediate statistics, including the general
properties of the bracket $\left[ u,v\right] _{n}$, relations of creation
and annihilation operators of intermediate statistics, and a different
construction of the number operator of intermediate statistics.

Coherent states in our cases are just the eigenstates of the annihilation
operator. Bosonic and fermionic coherent states have been widely discussed
\cite{KS}. We will construct the eigenstates of the annihilation operator of
intermediate statistics, the intermediate-statistics coherent states. The
result shows that the construction of the intermediate-statistics coherent
state is not unique.

Bosonic and fermionic oscillators play important roles in many physical
theories. In this paper we construct a kind of intermediate-statistics
oscillator which returns to bosonic oscillators when $n\rightarrow\infty$
and returns to fermionic oscillators when $n=1$, and calculate the energy
spectrum of such a system.

Ref. \cite{Ours2} shows that by use of only a single set of creation and
annihilation operators of intermediate statistics, one can establish a kind
of representation of the angular momentum ($su(2)$) algebra. Notice that a
bosonic realization of the $su(2)$ algebra needs two independent sets of
bosonic operators (the Schwinger representation) \cite{Schwinger};
otherwise, the realization with only one set of operators needs the help of
some kinds of intermediate statistics (e.g., the Holstein--Primakoff
representation) \cite{HP}. The representation of $su(2)$ algebra has been
discussed by many authors \cite{TJ}. In this paper we give a more general
discussion of the intermediate-statistics representation of $su(2)$ algebra
and present some kinds of representations of $su(2)$ algebra.

In this paper we (1) give a general discussion of the properties of the
intermediate-statistics quantum bracket (Sec. \ref{II} and Appendix A) and
the operator relations of intermediate statistics (Appendix B), (2)
construct a kind of intermediate-statistics coherent state (Sec. \ref{III}),
(3) construct an intermediate-statistics oscillator and calculate its
spectrum (Sec. \ref{IV}), and (4) present some realizations of angular
momentum ($su(2)$) algebra based on creation and annihilation operators of
intermediate statistics (Sec. \ref{V}). The conclusions are summarized in
Sec. \ref{VI}.

\section{Properties of the intermediate-statistics quantum bracket\label{II}}

In this section, we will present some general results of the
intermediate-statistics quantum bracket, $\left[ u,v\right] _{n}$. In the
following, $u$, $v$, $w$, and $o$ denote operators, and $\lambda$ denotes a
c-number.

The intermediate-statistics quantum bracket of an operator and itself is%
\begin{equation}
\left[ u,u\right] _{n}=\left( 1-e^{i\theta_{n}}\right) u^{2},
\end{equation}
and the intermediate-statistics quantum bracket of an operator and a
c-number is
\begin{equation}
\left[ u,\lambda\right] _{n}=\left[ \lambda,u\right] _{n}=\left(
1-e^{i\theta_{n}}\right) \lambda u.
\end{equation}
Some basic properties of the intermediate-statistics quantum bracket are as
follows:%
\begin{align}
\left[ u\pm v,w\right] _{n} & =\left[ u,w\right] _{n}\pm\left[ v,w\right]
_{n},  \notag \\
\left[ w,u\pm v\right] _{n} & =\left[ w,u\right] _{n}\pm\left[ w,v\right]
_{n},  \notag \\
\left[ u,\lambda v\right] _{n} & =\left[ \lambda u,v\right] _{n}=\lambda%
\left[ u,v\right] _{n},  \notag \\
\left[ u,v\right] _{n} & =-e^{-i\theta_{n}}\left[ v,u\right]
_{n}-2i\sin\theta_{n}vu.
\end{align}
Notice that in the case of commutation ($n\rightarrow\infty$) or
anticommutation ($n=1$) the phase factor $e^{i\theta_{n}}$ is $1$ or $-1$.
The relations between the intermediate-statistics quantum bracket and the
commutator and anticommutator are%
\begin{align}
\left[ u,v\right] _{n}-\left[ v,u\right] _{n} & =\left( 1+e^{i\theta
_{n}}\right) \left[ u,v\right] ,  \notag \\
\left[ u,v\right] _{n}+\left[ v,u\right] _{n} & =\left( 1-e^{i\theta
_{n}}\right) \left\{ u,v\right\} .
\end{align}

A general result of the intermediate-statistics quantum bracket of the
product of an arbitrary number of operators is
\begin{align}
\left[ u_{1}\cdots u_{k},v_{1}\cdots v_{l}\right] _{n} & =\overset{k}{%
\underset{i=1}{\sum}}\overset{l}{\underset{j=1}{\sum}}u_{1}\cdots
u_{i-1}v_{1}\cdots v_{j-1}\left[ u_{i},v_{j}\right] v_{j+1}\cdots
v_{l}u_{i+1}\cdots u_{k}  \notag \\
& +\left( 1-e^{i\theta_{n}}\right) v_{1}\cdots v_{l}u_{1}\cdots u_{k}.
\label{5}
\end{align}
Moreover, the properties of the two-fold intermediate-statistics quantum
bracket are
\begin{align}
& \left[ \left[ u,v\right] _{n},w\right] _{n}+\left[ \left[ w,u\right] _{n},v%
\right] _{n}+\left[ \left[ v,w\right] _{n},u\right] _{n}+\left[ \left[ v,u%
\right] _{n},w\right] _{n}+\left[ \left[ w,v\right] _{n},u\right] _{n}+\left[
\left[ u,w\right] _{n},v\right] _{n}  \notag \\
& =\left( 1-e^{i\theta_{n}}\right) ^{2}\left( uvw+wuv+vwu+vuw+wvu+uwv\right)
,  \label{6a} \\
& \left[ \left[ u,v\right] _{n},w\right] _{n}+\left[ \left[ w,u\right] _{n},v%
\right] _{n}+\left[ \left[ v,w\right] _{n},u\right] _{n}-\left[ \left[ v,u%
\right] _{n},w\right] _{n}-\left[ \left[ w,v\right] _{n},u\right] _{n}-\left[
\left[ u,w\right] _{n},v\right] _{n}  \notag \\
& =\left( 1-e^{i2\theta_{n}}\right) \left( uvw+wuv+vwu-vuw-wvu-uwv\right)
\label{6b}
\end{align}
Eq. (\ref{6b}) is in fact a generalized Jacobi identity. Generalizing these
relations to the case of $k$ operators, we have%
\begin{equation}
\underset{p}{\sum}[{\cdots\lbrack\lbrack u_{1},u_{2}]_{n},u_{3}]}_{n},{%
\cdots,u_{k}}]_{n}=\left( 1-e^{i\theta_{n}}\right) ^{k-1}\underset{p}{\sum}%
u_{1}u_{2}\cdots u_{k}.
\end{equation}
In this equation and following, by $\underset{p}{\sum}$ we mean the sum over
all permutations of the operators, and $\underset{r}{\sum}$ over all cyclic
permutations. Furthermore, we also have the following results:%
\begin{align}
\underset{p}{\sum}\left[ u_{1}\cdots u_{i},u_{i+1}\cdots u_{k}\right] _{n} &
=\left( 1-e^{i\theta_{n}}\right) \underset{p}{\sum}u_{1}\cdots u_{k},
\label{57} \\
\underset{r}{\sum}\left[ u_{1}\cdots u_{i},u_{i+1}\cdots u_{k}\right] _{n} &
=\left( 1-e^{i\theta_{n}}\right) \underset{r}{\sum}u_{1}\cdots u_{k}.
\label{58}
\end{align}
It should be pointed out that wherever the comma in the
intermediate-statistics quantum brackets in Eqs. (\ref{57}) and (\ref{58})
appears, the right-hand sides are the same, i.e.,%
\begin{align}
\underset{p}{\sum}\left[ u_{1}\cdots u_{k-1},u_{k}\right] _{n} & =\underset{p%
}{\sum}\left[ u_{1}\cdots u_{k-2},u_{k-1}u_{k}\right] _{n}=\cdots=\underset{p%
}{\sum}\left[ u_{1},u_{2}\cdots u_{k}\right] _{n}, \\
\underset{r}{\sum}\left[ u_{1}\cdots u_{k-1},u_{k}\right] _{n} & =\underset{r%
}{\sum}\left[ u_{1}\cdots u_{k-2},u_{k-1}u_{k}\right] _{n}=\cdots=\underset{r%
}{\sum}\left[ u_{1},u_{2}\cdots u_{k}\right] _{n}.
\end{align}

More properties of the intermediate-statistics quantum brackets are listed
in Appendix A.

\section{Intermediate-statistics coherent state\label{III}}

In this section we introduce the intermediate-statistics coherent state. We
will show that the construction of the intermediate-statistics coherent
state is not unique.

The concept of coherent states is applied to a wide class of objects \cite%
{KS}. The coherent state in this case is the eigenstate of the annihilation
operator. The coherent states in the Bose case and in the Fermi case are
very different due to the difference between the exchange symmetries of
bosons and fermions. For constructing the fermionic coherent state, one
needs to use the Grassmann number --- anticommuting c-numbers. The
annihilation operator corresponding to intermediate statistics is of course
different from the Bose and the Fermi cases; for constructing such a kind of
the coherent state we need to introduce the generalized Grassmann number.

Let $\left\vert \psi \right\rangle $ be the eigenstate of the annihilation
operator $b$:%
\begin{equation}
b\left\vert \psi \right\rangle =\psi \left\vert \psi \right\rangle ,
\label{e040}
\end{equation}%
where the state $\left\vert \psi \right\rangle $ is the
intermediate-statistics coherent state. For constructing the coherent state $%
\left\vert \psi \right\rangle $, we assume%
\begin{equation}
\left\vert \psi \right\rangle =M\left[ \left\vert 0\right\rangle _{n}+\delta
\left( 1,n\right) \left\vert 1\right\rangle _{n}\psi +\delta \left(
2,n\right) \left\vert 2\right\rangle _{n}\psi ^{2}+\cdots \delta \left(
n,n\right) \left\vert n\right\rangle _{n}\psi ^{n}\right] ,  \label{coherent}
\end{equation}%
where $M$ is the normalization constant, and $\delta \left( i,n\right) $ $%
\left( i=1,2,\cdots ,n\right) $ are coefficients to be determined. For
satisfying Eq. (\ref{e040}), one has to assume that $\psi $ here is neither
an ordinary commuting c-number nor an anticommuting c-number like that in
the Fermi case. $\psi $ must be a generalized Grassmann number satisfying
\begin{equation}
\psi ^{n+1}=0.
\end{equation}%
When $n=1$ we have $\psi ^{2}=0$ and $\psi $ returns to the Grassmann
number. Like that in the Fermi case \cite{KS}, $\psi \left\vert \nu
\right\rangle _{n}\neq \left\vert \nu \right\rangle _{n}\psi $. Assuming that

\begin{equation}
\psi\left\vert \nu\right\rangle _{n}=\lambda\left( \nu,n\right) \left\vert
\nu\right\rangle _{n}\psi,
\end{equation}
one can check that the coefficients in Eq. (\ref{coherent}) can be taken as

\begin{equation}
\delta\left( \nu,n\right) =\overset{\nu-1}{\underset{j=0}{\prod}}\frac{%
\lambda\left( j,n\right) }{\sqrt{\left\langle j+1\right\rangle _{n}}},
\label{e}
\end{equation}
where $\langle\nu\rangle_{n}=\left[ 1-e^{i2\pi\nu/\left( n+1\right) }\right]
/\left[ 1-e^{i2\pi/\left( n+1\right) }\right] =\sum_{j=0}^{\nu-1}e^{i2\pi
j/\left( n+1\right) }$, and the relations $a^{\dag}|\nu\rangle_{n}=\sqrt{%
\left\langle \nu+1\right\rangle _{n}}|\nu+1\rangle_{n}$ and $%
b|\nu\rangle_{n}=\sqrt{\left\langle \nu\right\rangle _{n}}|\nu -1\rangle_{n}$
\cite{Ours2} have been used. Eq. (\ref{e}) gives the relations between $%
\delta\left( \nu,n\right) $ and $\lambda\left( \nu,n\right) $. The
normalization constant, $M$, is determined by $\left\langle \psi
|\psi\right\rangle =1$, where the adjoint state vector%
\begin{equation}
\left\langle \psi\right\vert =M\left[ \left\langle 0\right\vert
_{n}+\delta\left( 1,n\right) ^{\ast}\bar{\psi}\left\langle 1\right\vert
_{n}+\delta\left( 2,n\right) ^{\ast}\bar{\psi}^{2}\left\langle 2\right\vert
_{n}+\cdots\delta\left( n,n\right) ^{\ast}\bar{\psi}^{n}\left\langle
n\right\vert _{n}\right] .
\end{equation}
Then we have%
\begin{equation}
M=\left[ 1+\underset{m=1}{\overset{n}{\sum}}\left( \bar{\psi}\psi\right)
^{m}\left\vert \delta\left( m,n\right) \right\vert ^{2}\right] ^{-1/2}.
\end{equation}
As in the Fermi case, $\bar{\psi}$ is independent of $\psi$ and is not a
proper mathematical adjoint \cite{KS}.

There is not only one way to construct $\lambda\left( \nu,n\right) $. $%
\lambda\left( \nu,n\right) $, for example, can be taken as%
\begin{equation}
\lambda\left( \nu,n\right) =e^{\pm i2\pi\nu/\left( n+1\right) },
\end{equation}
in this case%
\begin{equation}
\delta\left( \nu,n\right) =\frac{e^{\pm i\pi\nu\left( \nu-1\right) /\left(
n+1\right) }\left[ 1-e^{i2\pi/\left( n+1\right) }\right] ^{\nu/2}}{%
\prod\nolimits_{j=1}^{\nu}\sqrt{1-e^{i2\pi j/(n+1)}}};
\end{equation}
or%
\begin{equation}
\lambda\left( \nu,n\right) =\left( -1\right) ^{\nu},
\end{equation}
in this case%
\begin{equation}
\delta\left( \nu,n\right) =\frac{\left( -1\right) ^{\left( \nu-1\right)
\nu/2}\left[ 1-e^{i2\pi/(n+1)}\right] ^{\nu/2}}{\prod\nolimits_{j=1}^{\nu }%
\sqrt{1-e^{i2\pi j/(n+1)}}}.
\end{equation}
Especially, in the Fermi case, i.e., $n=1$, the coherent state can be
constructed as $\left\vert \psi\right\rangle =M\left( \left\vert
0\right\rangle +\left\vert 1\right\rangle \psi\right) $. It can be checked
directly that the above constructions will return to the Fermi case when $%
n=1 $.

Moreover, we also have
\begin{equation}
\psi\left( b^{\dag}\right) ^{\nu}=\frac{\lambda\left( \nu,n\right) }{%
\lambda\left( 0,n\right) }\left( b^{\dag}\right) ^{\nu}\psi,\text{ }%
\psi\left( a^{\dag}\right) ^{\nu}=\frac{\lambda\left( \nu,n\right) }{%
\lambda\left( 0,n\right) }\left( a^{\dag}\right) ^{\nu}\psi,\text{ }%
b^{\nu}\psi=\frac{\lambda\left( \nu,n\right) }{\lambda\left( 0,n\right) }%
\psi b^{\nu},\text{\ }a^{\nu}\psi=\frac{\lambda\left( \nu,n\right) }{%
\lambda\left( 0,n\right) }\psi a^{\nu}.
\end{equation}

\section{Intermediate-statistics oscillator\label{IV}}

The Hamiltonians for bosonic and fermionic oscillators can be expressed as%
\begin{align}
H_{\mathrm{Bose}}& =\frac{1}{2}\left( a_{b}^{\dag }a_{b}+a_{b}a_{b}^{\dag
}\right) =a_{b}^{\dag }a_{b}+\frac{1}{2},  \label{BO} \\
H_{\mathrm{Fermi}}& =\frac{1}{2}\left( a_{f}^{\dag }a_{f}-a_{f}a_{f}^{\dag
}\right) =a_{f}^{\dag }a_{f}-\frac{1}{2},  \label{FO}
\end{align}%
where $a_{b}^{\dag }$, $a_{b}$ and $a_{f}^{\dag }$, $a_{f}$ are creation and
annihilation operators of bosons and fermions, respectively, obeying $%
[a_{b},a_{b}^{\dag }]=1$ and $\{a_{f},a_{f}^{\dag }\}=1$. As a
generalization of bosonic and fermionic oscillators, we can construct an
intermediate-statistics oscillator using the creation and the annihilation
operators of intermediate statistics, $a^{\dag }$ and $b$. The Hamiltonian
and the spectrum of such an intermediate-statistics oscillator should return
to the bosonic oscillator when $n\rightarrow \infty $ and return to the
fermionic oscillator when $n=1$.

The Hamiltonian of the intermediate-statistics oscillator should be a
quadratic form of creation and annihilation operators. It can be constructed
in the following general form:%
\begin{equation}
H=\frac{1}{4}\left[ \alpha\left( n\right) a^{\dag}b+\beta\left( n\right)
ba^{\dag}+h.c.\right] .
\end{equation}
In the two limit cases, $n=\infty$ and $n=1$, one has $a=b$, and then the
creation and the annihilation operators are Hermitian conjugate of each
other \cite{Ours2}. The coefficients $\alpha\left( n\right) $ and $%
\beta\left( n\right) $ should satisfy $\text{Re}\alpha\left( \infty\right)
=1 $, $\text{Re}\beta\left( \infty\right) =1$, $\text{Re}\alpha\left(
1\right) =1$, and $\text{Re}\beta\left( 1\right) =-1$ for recovering the
Bose case, Eq. (\ref{BO}), when $n=\infty$, and the Fermi case, Eq. (\ref{FO}%
), when $n=1$, respectively. The choice of $\alpha\left( n\right) $ and $%
\beta\left( n\right) $ is not unique. One of the simplest choices is $%
\alpha\left( n\right) =1$ and $\beta\left( n\right) =e^{-i2\pi/\left(
n+1\right) }$, i.e.,

\begin{equation}
H=\frac{1}{4}\left[ a^{\dag}b+e^{-i2\pi/\left( n+1\right) }ba^{\dag
}+b^{\dag}a+e^{i2\pi/\left( n+1\right) }ab^{\dag}\right] .   \label{H}
\end{equation}

In this case, we have%
\begin{align}
\left[ H,a^{\dagger}\right] & =\cos\frac{2\left( N-1\right) \pi}{n+1}%
a^{\dagger}=a^{\dagger}\cos\frac{2N\pi}{n+1},\text{ }\left[ H,a\right] =-\cos%
\frac{2N\pi}{n+1}a=-a\cos\frac{2\left( N-1\right) \pi}{n+1},  \notag \\
\left[ H,b^{\dagger}\right] & =\cos\frac{2\left( N-1\right) \pi}{n+1}%
b^{\dagger}=b^{\dagger}\cos\frac{2N\pi}{n+1},\text{ }\left[ H,b\right] =-\cos%
\frac{2N\pi}{n+1}b=-b\cos\frac{2\left( N-1\right) \pi}{n+1}.
\end{align}

The spectrum should be discussed separately in four cases: $n=4t+1$, $n=4t+2$%
, $n=4t+3$, and $n=4t+4$, where $t\geqslant0$ and $n\geq1$.

The case of $n=4t+1$. The spectrum is

\begin{equation}
E_{k}^{\left( n\right) }=\cos^{2}\frac{\pi}{n+1}+\frac{1}{2}\csc\frac{\pi }{%
n+1}\sin\frac{\left( 4k-n-1\right) \pi}{2\left( n+1\right) },\text{ }%
k=0,\cdots,\frac{1}{2}\left( n+1\right) .   \label{fermilike}
\end{equation}
Except the highest energy level which is non-degenerate, all the energy
levels are two-fold degenerate. The total number of energy levels is $\left(
n+3\right) /2$.

The energy of the ground state%
\begin{equation*}
E_{0}^{\left( n\right) }=\cos^{2}\frac{\pi}{n+1}-\frac{1}{2}\csc\frac{\pi }{%
n+1}.
\end{equation*}
The highest energy is%
\begin{equation*}
E_{\left( n+1\right) /2}^{\left( n\right) }=\cos^{2}\frac{\pi}{n+1}+\frac{1}{%
2}\csc\frac{\pi}{n+1}.
\end{equation*}

The case of $n=4t+2$. The spectrum is
\begin{equation}
E_{k}^{\left( n\right) }=\cos ^{2}\frac{\pi }{n+1}+\frac{1}{2}\csc \frac{\pi
}{n+1}\sin \frac{\left( 2k-n\right) \pi }{2\left( n+1\right) },\ k=0,\cdots
n.
\end{equation}%
All energy levels in this case are non-degenerate. The total number of
energy levels is $n+1$.

The energy of the ground state%
\begin{equation*}
E_{0}^{\left( n\right) }=\cos^{2}\frac{\pi}{n+1}-\frac{1}{2}\csc\frac{\pi }{%
n+1}\sin\frac{n\pi}{2\left( n+1\right) }.
\end{equation*}
The highest energy is%
\begin{equation*}
E_{n}^{\left( n\right) }=\cos^{2}\frac{\pi}{n+1}+\frac{1}{2}\csc\frac{\pi }{%
n+1}\sin\frac{n\pi}{2\left( n+1\right) }.
\end{equation*}

The case of $n=4t+3$. The spectrum is
\begin{equation}
E_{k}^{\left( n\right) }=\cos ^{2}\frac{\pi }{n+1}+\frac{1}{2}\csc \frac{\pi
}{n+1}\sin \frac{\left( 4k-n+1\right) \pi }{2\left( n+1\right) },\
k=0,1,\cdots ,\frac{1}{2}\left( n-1\right) .
\end{equation}%
All energy levels are two-fold degenerate. The total number of energy levels
is $\left( n+1\right) /2$.

The energy of the ground state%
\begin{equation*}
E_{0}^{\left( n\right) }=\cos^{2}\frac{\pi}{n+1}-\frac{1}{2}\csc\frac{\pi }{%
n+1}\sin\frac{\left( n-1\right) \pi}{2\left( n+1\right) }.
\end{equation*}
The highest energy is%
\begin{equation*}
E_{\left( n-1\right) /2}^{\left( n\right) }=\cos^{2}\frac{\pi}{n+1}+\frac{1}{%
2}\csc\frac{\pi}{n+1}\sin\frac{\left( n-1\right) \pi}{2\left( n+1\right) }.
\end{equation*}

The case of $n=4t+4$ is just the same as the case of $n=4t+2$.

The above result shows that, for an intermediate-statistics oscillator with
a finite $n$, the total number of energy levels is finite, and the energy
levels are often degenerate.

The intermediate-statistics oscillator will return to bosonic and fermionic
oscillators when $n\rightarrow \infty $ and $n=1$ as follows.

(a) $n=1$: \textit{The Fermi case}. When $n=1$, the intermediate-statistics
oscillator will return to a fermionic oscillator. $n=1$ corresponds to the
case of $n=4t+1$. The spectrum, Eq. (\ref{fermilike}), reduces to

\begin{equation*}
E_{k}^{\left( 1\right) }=-\frac{1}{2}\cos k\pi,\ k=0,1.
\end{equation*}
The total number of energy levels is $\left( n+3\right) /2=2$. The energy of
the ground state%
\begin{equation*}
E_{0}^{\left( 1\right) }=-\frac{1}{2},
\end{equation*}
and the highest energy is%
\begin{equation*}
E_{1}^{\left( 1\right) }=\frac{1}{2}.
\end{equation*}
This is just a fermionic oscillator.

(b) $n\rightarrow \infty $: \textit{The Bose case}. When $n=\infty $, the
intermediate-statistics oscillator will return to a bosonic oscillator. In
Ref. \cite{Ours1} we have shown that, in the theory of statistical
mechanics, for recovering Bose-Einstein statistics from Gentile statistics,
besides the condition $n\rightarrow \infty $, one also needs an additional
condition $\lim_{n\rightarrow \infty ,N\rightarrow \infty }N/n=0$, where $N$
is the total number of particles in the system. Concretely, $%
\lim_{n\rightarrow \infty ,N\rightarrow \infty }N/n$ can also take other
nonzero values, and different nonzero values of such a limit correspond to
different kinds of intermediate statistics \cite{Ourspre}; only $%
\lim_{n\rightarrow \infty ,N\rightarrow \infty }N/n=0$ corresponds to the
Bose case. In our cases, $\nu $ is the occupation number of one quantum
state. Therefore, in the Bose case $\nu \ll n$. Thus, for recovering the
bosonic oscillator, the condition $\nu \ll n$ is needed.

The spectrum of the Hamiltonian Eq. (\ref{H}) can also be expressed as%
\begin{equation}
E^{\left( n\right) }\left( \nu\right) =\frac{1}{2}\csc\frac{\pi}{n+1}\left[
\sin\frac{\left( 2\nu-1\right) \pi}{n+1}+\sin\frac{2\pi}{n+1}\cos\frac{\pi}{%
n+1}\right] .
\end{equation}
When $\nu\ll n$, the spectrum%
\begin{equation*}
\left. E^{\left( n\right) }\left( \nu\right) \right\vert _{\nu\ll n}\approx%
\frac{\pi\nu}{n+1}\cot\frac{\pi}{n+1}+\frac{1}{2}\cos\frac{2\pi}{n+1}.
\end{equation*}
Taking $n\rightarrow\infty$, we have%
\begin{equation*}
E^{\left( \infty\right) }\left( \nu\right) =\lim_{n\rightarrow\infty }\left.
E^{\left( n\right) }\left( \nu\right) \right\vert _{\nu\ll n}=\nu+\frac{1}{2}%
.
\end{equation*}
This is just the spectrum of a bosonic oscillator. That is to say, in the
case of $\nu\ll n$, when $n\rightarrow\infty$ the intermediate-statistics
oscillator will return to a bosonic oscillator. Such a result agrees with
the conclusion drawn in Refs. \cite{Ours1} and \cite{Ourspre}: intermediate
statistics cannot recover the Bose case by only taking $n\rightarrow\infty$;
for recovering the Bose case, one also needs $\nu\ll n$.

\section{Intermediate-statistics representation of angular momentum ($su(2)$%
) algebra\label{V}}

It is shown in Ref. \cite{Ours2} that one cannot obtain a bosonic
realization of the angular momentum ($su(2)$) algebra by a single set of
bosonic creation and annihilation operators. Two kinds of representations of
angular momentum operators are the Schwinger representation \cite{Schwinger}
and the Holstein--Primakoff representation \cite{HP}, which are successful
in describing magnetism in various quantum systems \cite{SHP}. The Schwinger
representation needs two sets of independent boson operators, $a_{1}$ and $%
a_{2}$: $J_{+}=a_{1}^{\dag }a_{2}$, $J_{-}=a_{2}^{\dag }a_{1}$, and $J_{z}=%
\frac{1}{2}\left( a_{1}^{\dag }a_{1}-a_{2}^{\dag }a_{2}\right) $. The
Holstein-Primakoff representation, $J_{+}=\sqrt{2j-N}a$, $J_{-}=a^{\dag }%
\sqrt{2j-N}$, and $J_{z}=j-N$, where $N=0,1,\cdots ,2j$, though only useing
one set of creation and annihilation operators and $a$ and $a^{\dag }$
satisfying the bosonic commutation relations, is not a real bosonic
realization. This is because in the Holstein-Primakoff case the occupation
number $N$ is restricted to be no more than $2j$, but in Bose-Einstein
statistics $N$ can take any integer. The result given in Ref. \cite{Ours2}
shows that the angular momentum algebra can be represented in terms of a
single set of creation and annihilation operators in the case of
intermediate statistics.

There is a correspondence between the angular momentum and intermediate
statistics, which allows us to construct a representation in terms of
intermediate statistics. More concretely, when the maximum occupation number
is $n$, there are $n+1$ states: $|0\rangle_{n}$, $|1\rangle_{n}$,$\cdots$, $%
\left\vert n\right\rangle _{n}$; correspondingly, when the magnitude of the
angular momentum is $j=n/2$, there are also $2j+1=n+1$ states: $\left\vert
-j\right\rangle $, $\left\vert -j+1\right\rangle $,$\cdots$, $\left\vert
j\right\rangle $. In Ref. \cite{Ours2}, some special cases of the
realization of angular momentum ($su(2)$) algebra corresponding to
intermediate statistics, $j=1/2$, $1$, $3/2$, $2$, $5/2$, are discussed. In
this paper, we give a systematic analysis for the intermediate-statistics
realization of angular momentum algebra.

The angular momentum operators, $J_{+}$, $J_{-}$, and $J_{z}$ satisfy
\begin{align}
\left[ J_{+},J_{-}\right] & =2J_{z},  \label{87} \\
\left[ J_{z},J_{\pm}\right] & =\pm J_{\pm}.   \label{88}
\end{align}
Like in Ref. \cite{Ours2}, we represent $J_{z}$ by the particle number
operator:%
\begin{equation}
J_{z}=N-\dfrac{n}{2}.   \label{w}
\end{equation}
Thus we have%
\begin{equation}
J_{z}|\nu\rangle_{n}=\left( N-\dfrac{n}{2}\right) |\nu\rangle_{n}=\left( \nu-%
\dfrac{n}{2}\right) |\nu\rangle_{n}.
\end{equation}
For a given $n$, the range of value of $\nu$, the occupation number of one
single quantum state in intermediate statistics, is $0\leq\nu\leq n$, so the
values of $J_{z}$ are $-n/2,-n/2+1,\cdots,n/2$, a total of $n+1$,
corresponding to the $n+1$ components of the angular momentum $j=n/2$.

For satisfying Eqs. (\ref{87}) and (\ref{88}), notice that $\left[ N,J_{\pm }%
\right] =\pm J_{\pm }$, we can choose $J_{+}$ and $J_{-}$ satisfying%
\begin{align}
J_{+}\left\vert \nu \right\rangle _{n}& =c_{+}\left( \nu \right) \left\vert
\nu +1\right\rangle _{n},  \notag \\
J_{-}\left\vert \nu \right\rangle _{n}& =c_{-}\left( \nu \right) \left\vert
\nu -1\right\rangle _{n},  \label{Jpm}
\end{align}%
so that%
\begin{align}
\left[ J_{z},J_{\pm }\right] |\nu \rangle _{n}& =\left[ c_{\pm }\left( \nu
\right) \left( \nu \pm 1-\frac{n}{2}\right) -c_{\pm }\left( \nu \right)
\left( \nu -\frac{n}{2}\right) \right] |\nu \pm 1\rangle _{n}  \notag \\
& =\pm J_{\pm }|\nu \pm 1\rangle _{n},
\end{align}%
and then Eq. (\ref{88}) is automatically satisfied. The choice of $J_{\pm }$
satisfying Eq. (\ref{Jpm}) is not unique. In the following we will consider
some kinds of constructions.

One possible construction is%
\begin{align}
J_{+}& =\sum_{l}\lambda _{l}^{\ast }A^{l}a^{\dagger },  \notag \\
J_{-}& =\sum_{q}\lambda _{q}a\left( A^{\dagger }\right) ^{q},  \label{y}
\end{align}%
where the operator $A$ satisfies%
\begin{equation}
\lbrack A,N]=[A^{\dagger },N]=0.  \label{z}
\end{equation}%
Obviously, such a choice satisfies Eq. (\ref{88}). Eq. (\ref{z}) implies
that the operator $A$ can be chosen as
\begin{equation}
\text{ }A=N,\text{ }A=\text{ }a^{\dagger }b\text{ or }b^{\dagger }a,\text{ }%
A=a^{\dagger }a=b^{\dagger }b,\text{ }A=aa^{\dagger }=bb^{\dagger },\text{
etc}.
\end{equation}%
The coefficients can be determined by Eq. (\ref{87}). For example, when $%
A=a^{\dagger }b$, Eq. (\ref{87}) gives
\begin{equation}
\sum_{lq}\lambda _{l}^{\ast }\lambda _{q}\left[ \left\vert \left\langle \nu
\right\rangle _{n}\right\vert \left( \left\langle \nu \right\rangle
_{n}^{\ast }\right) ^{q}\left\langle \nu \right\rangle _{n}^{l}-\left\vert
\left\langle \nu +1\right\rangle _{n}\right\vert \left( \left\langle \nu
+1\right\rangle _{n}^{\ast }\right) ^{q}\left\langle \nu +1\right\rangle
_{n}^{l}\right] =2\nu -n.  \label{e010}
\end{equation}%
This is a set of $n+1$ equations. In principle, one can obtain a realization
of angular momentum ($su(2)$) algebra by solving the coefficients $\lambda
_{l}$ from this set of equations. One can prove that such a set of equations
always possess solutions. Eq. (\ref{e010}) is a set of linear equations of $%
\lambda _{l}^{\ast }\lambda _{q}$. However, once one wants to solve $\lambda
_{l}$, he will encounter high-order algebraic equations. When the order of
the equation is high, it is difficult to solve. Nevertheless, in a concrete
problem, in which $n$ is given, one can always obtain the solution, and then
obtain the realization of the angular momentum ($su(2)$) algebra.

Note that the realization given in Eq. (\ref{y}) is not unique. There are
still other choices, for example,%
\begin{align}
J_{+} & =\sum_{l}\lambda_{l}^{\ast}\left( A^{l}a^{\dagger}+B^{l}b^{\dagger
}\right) ,  \notag \\
J_{-} & =\sum_{q}\lambda_{q}\left[ a\left( A^{\dagger}\right) ^{q}+b\left(
B^{\dagger}\right) ^{q}\right] ,
\end{align}
where%
\begin{equation}
\lbrack B,N]=[B^{\dagger},N]=0.
\end{equation}
The coefficients $\lambda_{i}$ can be obtained by the same procedure given
above.

\section{Conclusions\label{VI}}

In this paper, we discuss the properties of the intermediate-statistics
quantum bracket, $\left[ u,v\right] _{n}=uv-e^{i\theta_{n}}vu$, which is
introduced in Ref. \cite{Ours2}. The operation $\left[ u,v\right] _{n}$ will
return to commutativity and anti-commutativity when $n\rightarrow\infty$ or $%
n=1$. The physical meaning of $n$ is clear. It denotes the maximum
occupation number of statistics: $n\rightarrow\infty$ and $n=1$ correspond
to Bose and Fermi cases, and the other values of $n$ correspond to
intermediate statistics. That is to say, if the commutator reflects the
properties of a bosonic system, and the anti-commutator reflects the
properties of a fermionic system, the intermediate-statistics quantum
bracket, $\left[ u,v\right] _{n}$, corresponds to the system obeying
intermediate statistics.

An operator realization of intermediate statistics is given in Ref. \cite%
{Ours2}. In this paper, we give a more detailed discussion of the operator
realization. Especially, some operator relations corresponding to
intermediate statistics are provided.

The coherent state is an important concept in physics. The bosonic and
fermionic coherent states have been discussed in many literatures \cite{KS}.
In this paper we construct the intermediate-statistics coherent state. The
analysis shows that the construction of the intermediate-statistics coherent
state is not unique. In this paper, we provide two constructions.

The Fermi oscillator and the Bose oscillator are very important models in
quantum mechanics. Based on the operator realization of intermediate
statistics, we construct a kind of intermediate-statistics oscillator which
returns to the bosonic oscillator when $n\rightarrow\infty$ ($\nu\ll n$),
and returns to the fermionic oscillator when $n=1$. Its energy spectrum is
calculated.

In this paper, we provide a more general discussion of the
intermediate-statistics representation of angular momentum algebra. Our
result shows that one can construct more than one representation of angular
momentum algebra by a single set of creation and annihilation operators.
Note that one cannot obtain a representation of angular momentum algebra by
a single set of bosonic operators.

\begin{acknowledgements}
The authors W.-S. D. and M. X. are very indebted to Dr G.
Zeitrauman for his encouragement. This work is supported in part
by NSF of China, under Project No.10605013 and No.10675088.
\end{acknowledgements}

\appendix

\section{Properties of the intermediate-statistics quantum bracket}

Eq. (\ref{5}) gives a very general result of the intermediate-statistics
quantum bracket. In practice the following special results are often useful:
\begin{align}
\left[ u_{1}\cdots u_{k},v\right] _{n}& =\underset{i=1}{\overset{k}{\sum }}%
u_{1}\cdots u_{i-1}\left[ u_{i},v\right] u_{i+1}\cdots u_{k}+\left(
1-e^{i\theta _{n}}\right) vu_{1}\cdots u_{k},  \label{21} \\
\left[ v,u_{1}\cdots u_{k}\right] _{n}& =\overset{k}{\underset{i=1}{\sum }}%
u_{1}\cdots u_{i-1}\left[ v,u_{i}\right] u_{i+1}\cdots u_{k}+\left(
1-e^{i\theta _{n}}\right) u_{1}\cdots u_{k}v.  \label{22}
\end{align}%
When $u_{1}=\cdots =u_{k}=u$ and $v_{1}=\cdots =v_{l}=v$, Eq. (\ref{5})
becomes
\begin{equation}
\left[ u^{k},v^{l}\right] _{n}=\overset{k}{\underset{i=1}{\sum }}\overset{l}{%
\underset{j=1}{\sum }}u^{i-1}v^{j-1}\left[ u,v\right] v^{l-j}u^{k-i}+\left(
1-e^{i\theta _{n}}\right) v^{l}u^{k}.  \label{41}
\end{equation}%
We also have
\begin{subequations}
\begin{equation*}
\left[ u^{k},v^{m}\right] _{n}=\frac{1}{\left( 1-e^{i\theta _{n}}\right)
^{k+m-2}}\underset{k-1}{[\underbrace{\left[ \cdots \left[ u,u\right]
_{n}\cdots ,u\right] _{n}}},\underset{m-1}{\underbrace{\left[ \cdots \left[
v,v\right] _{n}\cdots ,v\right] _{n}]_{n}}}
\end{equation*}%
and
\end{subequations}
\begin{subequations}
\begin{align*}
\left[ u^{k},v\right] _{n}+\left[ u^{k-1}v,u\right] _{n}& =\left[ u^{k-1},u%
\right] _{n}v+\left[ u^{k-1},v\right] _{n}u, \\
\left[ v,u^{k}\right] _{n}+\left[ u,vu^{k-1}\right] _{n}& =v\left[ u,u^{k-1}%
\right] _{n}+u\left[ v,u^{k-1}\right] _{n}.
\end{align*}%
\ \ Some useful special cases of Eqs. (\ref{21}) and (\ref{22}) are
\end{subequations}
\begin{subequations}
\begin{align*}
\left[ uv,w\right] _{n}& =u\left[ v,w\right] +\left[ u,w\right] v+\left(
1-e^{i\theta _{n}}\right) wuv, \\
\left[ w,uv\right] _{n}& =\left[ w,u\right] v+u\left[ w,v\right] +\left(
1-e^{i\theta _{n}}\right) uvw,
\end{align*}%
\end{subequations}
\begin{align*}
\left[ uvw,o\right] _{n}& =\left[ u,o\right] vw+u\left[ v,o\right] w+uv\left[
w,o\right] +\left( 1-e^{i\theta _{n}}\right) ouvw, \\
\left[ o,uvw\right] _{n}& =\left[ o,u\right] vw+u\left[ o,v\right] w+uv\left[
o,w\right] +\left( 1-e^{i\theta _{n}}\right) uvwo,
\end{align*}%
and
\begin{align*}
\left[ uv,wo\right] _{n}& =\frac{1}{1-e^{i2\theta _{n}}}\left( \left[ \left[
u,v\right] _{n},\left[ w,o\right] _{n}\right] _{n}+e^{i\theta _{n}}\left[ %
\left[ v,u\right] _{n},\left[ w,o\right] _{n}\right] _{n}\right.  \\
& \left. +e^{i\theta _{n}}\left[ \left[ u,v\right] _{n},\left[ o,w\right]
_{n}\right] _{n}+e^{i2\theta _{n}}\left[ \left[ v,u\right] _{n},\left[ o,w%
\right] _{n}\right] _{n}\right) , \\
\left[ uv,wo\right] _{n}& =u\left[ v,w\right] _{n}o+e^{i\theta _{n}}uw\left[
v,o\right] +e^{i\theta _{n}}\left[ u,w\right] ov+e^{i\theta _{n}}w\left[ u,o%
\right] v.
\end{align*}

Eqs. (\ref{5})-(\ref{58}) are very general results. We can also obtain the
corresponding properties for commutators or anticommutators by taking $%
n\rightarrow\infty$ or $n=1$, which are often useful.

\textit{Commutator case}:
\begin{equation*}
\left[ u_{1}\cdots u_{k},v_{1}\cdots v_{l}\right] =\overset{k}{\underset{i=1}%
{\sum}}\overset{l}{\underset{j=1}{\sum}}u_{1}\cdots u_{i-1}v_{1}\cdots
v_{j-1}\left[ u_{i},v_{j}\right] v_{j+1}\cdots v_{l}u_{i+1}\cdots u_{k},
\end{equation*}
and
\begin{align}
\overset{3}{\underset{i,j,k=1}{\sum}}\left\vert \varepsilon_{ijk}\right\vert %
\left[ \left[ u_{i},u_{j}\right] ,u_{k}\right] & =0,  \label{80} \\
\overset{3}{\underset{i,j,k=1}{\sum}}\varepsilon_{ijk}\left[ \left[
u_{i},u_{j}\right] ,u_{k}\right] & =0.   \label{81}
\end{align}
Eq. (\ref{81}) is just the Jacobi identity. Moreover,%
\begin{equation*}
\underset{p}{\sum}\left[ \left[ \cdots\left[ u_{1},u_{2}\right]
\cdots,u_{k-1}\right] ,u_{k}\right] =0,
\end{equation*}
and%
\begin{align*}
\underset{p}{\sum}\left[ u_{1}\cdots u_{i},u_{i+1}\cdots u_{k}\right] & =0,
\\
\underset{r}{\sum}\left[ u_{1}\cdots u_{i},u_{i+1}\cdots u_{k}\right] & =0.
\end{align*}

\textit{Anticommutator case}:%
\begin{equation*}
\left\{ u_{1}\cdots u_{k},v_{1}\cdots v_{l}\right\} =\overset{k}{\underset{%
i=1}{\sum}}\overset{l}{\underset{j=1}{\sum}}u_{1}\cdots u_{i-1}v_{1}\cdots
v_{j-1}\left[ u_{i},v_{j}\right] v_{j+1}\cdots v_{l}u_{i+1}\cdots
u_{k}+2v_{1}\cdots v_{l}u_{1}\cdots u_{k},
\end{equation*}%
\begin{align*}
\overset{3}{\underset{i,j,k=1}{\sum}}\left\vert \varepsilon_{ijk}\right\vert
\left\{ \left\{ u_{i},u_{j}\right\} ,u_{k}\right\} & =4\underset{l,h,m=1}{%
\overset{3}{\sum}}\left\vert \varepsilon_{lhm}\right\vert u_{l}u_{h}u_{m}, \\
\overset{3}{\underset{i,j,k=1}{\sum}}\varepsilon_{ijk}\left\{ \left\{
u_{i},u_{j}\right\} ,u_{k}\right\} & =0,
\end{align*}%
\begin{equation*}
\underset{p}{\sum}\left\{ \left\{ \cdots\left\{ u_{1},u_{2}\right\}
\cdots,u_{k-1}\right\} ,u_{k}\right\} =2^{k-1}\underset{p}{\sum}u_{1}\cdots
u_{k},
\end{equation*}
and%
\begin{align*}
\underset{p}{\sum}\left\{ u_{1}\cdots u_{i},u_{i+1}\cdots u_{k}\right\} & =2%
\underset{p}{\sum}u_{1}\cdots u_{k}. \\
\underset{r}{\sum}\left\{ u_{1}\cdots u_{i},u_{i+1}\cdots u_{k}\right\} & =2%
\underset{r}{\sum}u_{1}\cdots u_{k}.
\end{align*}

\section{Operator relations of intermediate statistics}

Ref. \cite{Ours2} presents an operator realization of intermediate
statistics. Concretely, Ref. \cite{Ours2} constructs a set of creation,
annihilation, and number operators for intermediate statistics, Bose and
Fermi cases becoming its two limiting cases. In this appendix we will first
construct a realization for the number operator of intermediate statistics.
Then, using the properties of intermediate-statistics quantum brackets given
in the above section, we will give some operator relations for creation,
annihilation, and number operators of intermediate statistics.

For completeness, we rewrite the basic operator relations of intermediate
statistics given in Ref. \cite{Ours2}:%
\begin{equation}
\left[ b,a^{\dagger}\right] _{n}=1,\text{ }\left[ N,a^{\dagger}\right]
=a^{\dagger},\text{ }\left[ N,b\right] =-b,   \label{abN}
\end{equation}
where $a^{\dagger}$, $b$, and $N$ are creation, annihilation, and number
operators in intermediate statistics, respectively.

Ref. \cite{Ours2} gives a construction for the number operator $N$. However,
the construction of $N$ is not unique; it can also be constructed as%
\begin{equation}
N=\frac{n+1}{2\pi}\arcsin\left[ \frac{i}{2}\left( a^{\dagger}b-b^{\dag
}a+ab^{\dag}-ba^{\dagger}\right) \right] .   \label{N2}
\end{equation}
It can be checked directly that this construction satisfies Eq. (\ref{abN}).

Moreover, we give some operator relations of $a^{\dagger }$, $b$, and $N$:
\begin{subequations}
\begin{align*}
\left[ N,a^{\dagger }\right] _{n}& =\left[ \left( 1-e^{i\theta _{n}}\right)
N+e^{i\theta _{n}}\right] a^{\dagger },\text{ \ \ }\left[ a^{\dagger },N%
\right] _{n}=\left[ \left( 1-e^{i\theta _{n}}\right) N-1\right] a^{\dagger },
\\
\left[ N,b\right] _{n}& =\left[ \left( 1-e^{i\theta _{n}}\right)
N-e^{i\theta _{n}}\right] b,\text{ \ \ }\left[ b,N\right] _{n}=\left[ \left(
1-e^{i\theta _{n}}\right) N+1\right] b,
\end{align*}%
\end{subequations}
\begin{equation*}
\left[ Nb,a^{\dagger }b\right] =e^{i2\pi (N-1)/\left( n+1\right) }Nb\text{.}
\end{equation*}%
We also have
\begin{align*}
\lbrack \left( a^{\dagger }\right) ^{k}b,a^{\dagger }]_{n}& =[b\left(
a^{\dagger }\right) ^{k},a^{\dagger }]_{n}=\left( a^{\dagger }\right) ^{k},
\\
\lbrack b,a^{\dagger }b^{k}]_{n}& =[b,b^{k}a^{\dagger }]_{n}=b^{k}, \\
\left[ a^{\dagger }b^{2},a^{\dagger }\right] _{n}& =[b,\left( a^{\dagger
}\right) ^{2}b]_{n}=\left( 1+e^{i\theta _{n}}\right) a^{\dagger }b.
\end{align*}

Like the fact that intermediate statistics returns to Bose-Einstein and
Fermi-Dirac statistics when $n\rightarrow\infty$ and $n=1$, the
intermediate-statistics quantum bracket will return to commutator and
anticommutator in these two limiting cases. Therefore, when $n\rightarrow
\infty$ and $n=1$ the above results lead to the operator relations of
creation and annihilation operators in Bose-Einstein and Fermi-Dirac cases.

\end{document}